# A LATTICE CALCULATION

# OF THE

# HEAVY QUARK UNIVERSAL FORM FACTOR

*Jeffrey E. Mandula*
*U.S. Department of Energy*
*Division of High Energy Physics*
*Washington, DC  20585, USA*

*and*

*Michael C. Ogilvie*
*Department of Physics*
*Washington University*
*St. Louis, MO  63130, USA*

## ABSTRACT

A preliminary computation of the Isgur-Wise universal form factor using a lattice formulation of the Heavy Quark Effective Theory (HQET) is described and compared with the recent data from ARGUS and CLEO on $B \to D^* l \nu$ decay.

This letter describes a lattice computation of the Isgur-Wise function, also called the heavy quark universal form factor. The calculation uses the lattice implementation of the heavy quark effective theory developed recently by the authors[1]. The calculation is formulated and carried out entirely within the Isgur-Wise limit, in which the masses of heavy quarks are treated as infinite. We show the first lattice simulation results based on this formulation, and compare them to the recent data from ARGUS[2] and CLEO[3] on $B \to D^* l \nu$ decay, the actual process that comes closest to realizing the Isgur-Wise limit. Two other lattice calculations of this decay form factor have recently been published[4,5]. They treat the heavy quarks as Wilson fermions with a small hopping constant, but do not implement the heavy quark limit.

We briefly review the lattice formulation of the heavy quark effective theory. After Wick rotation to Euclidean space, the classical velocity, which is only a parameter, is complex when expressed in terms of Euclidean four-momenta:

$$v = (iv_0, \vec{v})$$
$$v_0 = \sqrt{1 + \vec{v}^2}$$
(1)

The reduced propagator for the heavy quark satisfies the Euclidean space equation

$$-iv \cdot D \tilde{S}^{(v)}(x) = -iv \cdot [\partial - igA(x)] \tilde{S}^{(v)}(x) = \delta(x)$$
(2)

with the boundary condition that $\tilde{S}^{(v)}(x)$ vanishes for negative $x_4$. In the presence of gauge interactions, the derivative is replaced by a gauge covariant derivative. In order to simply incorporate the boundary condition on the lattice, we represent the time derivative in the reduced

Dirac equation by means of an asymmetrical forward difference, but use a symmetrical centered difference for the space derivative. The lattice equation for the reduced propagator is:

$$v_0 [ U_4(x,x+\hat{t}) \tilde{S}^{(v)}(x+\hat{t},y) - \tilde{S}^{(v)}(x,y) ]$$
$$+ \sum_{\mu=1}^{3} \frac{-iv_\mu}{2} [U_\mu(x,x+\hat{\mu}) \tilde{S}^{(v)}(x+\hat{\mu},y) - U_\mu(x,x-\hat{\mu}) \tilde{S}^{(v)}(x-\hat{\mu},y) ] = \delta(x,y) \quad (3)$$

This is solved by simple forward recursion; no iterative techniques are needed. A symmetric time derivative could also be used, at an increase in computation.

Even for non-zero $\vec{v}$, only trivial computation is needed to construct the heavy quark propagator. This makes it entirely feasible to compute three-point functions, and a careful choice of heavy-light particle wave functions will improve results as in other lattice calculations.

The lattice heavy quark theory has some unfamiliar features which are worth noting. The symmetrical first difference results in fermion doubling. However, unlike the usual fermion situation, the contributions of the secondary modes automatically vanish in the zero spacing limit, without the necessity of adding a Wilson-like term. The reason is that they always occur in association with either a light Wilson quark or a gluon whose momentum is at the edge of the Brillouin zone, and those modes have energies on the order of the inverse lattice spacing. This may be seen in perturbation theory[6]. Note that to leading order in the inverse heavy quark mass, there are no closed heavy quark loops.

Another feature of the lattice heavy quark theory is that for residual momenta oriented oppositely to the classical velocity, the propagator grows with Euclidean time. The reason is that the falloff is governed by the difference between the actual energy of a mode and that of the zero-momentum mode, which is

$$\lim_{M \to \infty} \sqrt{M^2 + (M\vec{v} + \vec{p})^2} - M\sqrt{1 + \vec{v}^2} = \frac{\vec{v} \cdot \vec{p}}{v_0} \tag{4}$$

in the heavy quark limit. When this is negative, the "falloff" becomes growth. Conversely, at fixed Euclidean time, the heavy quark propagator grows in some directions in momentum space. The modes of the light quanta always compensate for this growth in simulations. In perturbation theory the contour in the Euclidean energy plane is determined by the boundary condition that heavy quarks propagate only forwards in time. The use of light staggered fermions would be problematic, since they have poles near the edge of the Brillouin zone.

The paradigmatic heavy quark process is $B \to D^* l \nu$, the weak decay of one meson containing a heavy quark into another. In the heavy quark limit, there is only one form factor that describes $B$ meson decay into both $D$ and $D^*$ mesons, and it is a function only of the classical velocities[7,8],

$$B \to D l \nu \quad , \quad D^* l \nu$$
$$\langle D v' | \bar{c} \gamma_\mu (1 - i\gamma_5) b | B v \rangle \sim \xi(v \cdot v') \tag{5}$$

All the dynamical information is contained in the function $\xi$, known as the Isgur-Wise function. It is a function only of the classical velocities, and is normalized to unity for forward decay because it also describes the conserved vector current taken between states of identical heavy mesons.

$$\xi(1) = 1 \tag{6}$$

For the process $B \to D^* l \nu$ the structure of the matrix element is

$$\langle D^* v', \varepsilon | \bar{c} \gamma_\mu (1 - i\gamma_5) b | B v \rangle = \sqrt{M_B M_{D^*}} \; C_{cb} \, \xi(v \cdot v') \, \Gamma_\mu(v, v', \varepsilon) \tag{7}$$

where $\Gamma_\mu$ is a kinematic trace of Dirac matrices and projectors, and the normalization constant $C_{cb}$ comes from the renormalization point dependent matching of the heavy quark effective field theory to QCD with quark masses much larger than the renormalization scale. It has been calculated to leading log approximation[9]:

$$C_{cb} = \left[ \frac{\alpha_s(M_{D^*})}{\alpha_s(M_B)} \right]^{6/(33 - 2N_b)} \left[ \frac{\alpha_s(M_B)}{\alpha_s(\mu)} \right]^{a(v \cdot v')} \tag{8}$$

where $N_b = 4$ is the number of "light" flavors below the heavier heavy ($B$) meson, and

$$a(v \cdot v') = \frac{8 [v \cdot v' \, r(v \cdot v') - 1]}{33 - 2N_c} \tag{9}$$

$N_c = 3$ is the number of "light" flavors below the lighter heavy ($D$) meson, and

$$r(v \cdot v') = \frac{\ln\left(v \cdot v' + \sqrt{(v \cdot v')^2 - 1}\right)}{\sqrt{(v \cdot v')^2 - 1}} \tag{10}$$

The renormalization scale μ dependence of the second factor in Eq. (8) is compensated by a scale dependence in ξ, in just such a way so that physical amplitudes do not depend on μ.

The amplitude for $B \to D^* l \nu$ is contained in the reduced three-point function

$$\tilde{G}^{(v_B, v_D)}(x_0, z_0, y_0) \sim \left\langle Tr\, s(y,x)\, \Gamma_D\, \tilde{S}^{(v_D)}(x,z)\, \Gamma_J\, \tilde{S}^{(v_B)}(z,y)\, \Gamma_B \right\rangle \tag{11}$$

The quark flow in this process is as shown in Figure 1. The residual momenta are eliminated by summing over all lattice sites on the $x_0$ and $z_0$ time slices. The Isgur-Wise function may be extracted from the three-point correlation by dividing out the normalized single meson propagators.

$$\xi(v \cdot v') = \lim_{x_0 \ll z_0 \ll z_0} \frac{Z_B^{1/2} Z_D^{1/2}\, \tilde{G}^{(v_B, v_D)}(x_0, z_0, y_0)}{\tilde{G}^{(v_B)}(x_0, z_0)\, \tilde{G}^{(v_D)}(z_0, y_0)} \tag{12}$$

The wave function renormalization constants are computed from the normalization of the asymptotic heavy-light meson propagators. Their evaluation, and the attendant numerical errors, can be avoided by taking advantage of the normalization of the Isgur-Wise function in the forward direction. We use three-point functions with both classical velocities set equal to $v$ and to $v'$ as normalization, and the expression from which we extract the universal form factor is

$$|\xi(v \cdot v')|^2 = \lim_{x_0 \gg z_0 \gg y_0} \frac{\tilde{G}^{(v,v')}(x_0, y_0, z_0)\, \tilde{G}^{(v',v)}(x_0, y_0, z_0)}{\tilde{G}^{(v,v)}(x_0, y_0, z_0)\, \tilde{G}^{(v',v')}(x_0, y_0, z_0)} \tag{13}$$

This form has the advantage that the overall normalization of the three-point function drops out completely. Only those renormalizations that depend on the classical velocity need be included.

We evaluated the three-point functions using the lattices made publicly available by Bernard and Soni[10]. There were an ensemble of 16 lattices of size $16^3 \times 24$ with lattice coupling $\beta = 5.7$. The light quarks were represented by Wilson quark propagators with hopping constant $\kappa = .164$. The heavy quark propagators were evaluated for four values of each component of the initial and final quark classical velocity: 0, .25, .50, and .75. That is, there were $4^6 = 4096$ combinations of $v_B$ and $v_D$. Of course, the many that were lattice rotations of each other were averaged to improve statistics, but even so there were more than a hundred different values of $v_B \cdot v_D$, lying in the interval [1,1.822].

To improve the overlap of the meson wave function with the ground state, we applied multiple single-link smearing steps. The light quark propagators were evaluated with point sources, and we applied a "Gaussian" smearing of the form

$$(1 + \alpha H)^n \tag{14}$$

to the heavy quark sources. H s a symmetrical single link displacement operator.

$$H(y,x) \equiv \sum_{i=1}^{3} [U(x+\hat{i},x)\delta_{y,x+\hat{i}} + U(x-\hat{i},x)\delta_{y,x-\hat{i}}] \tag{15}$$

The parameters used were n = 0, 5, 10 and $\alpha = 4.0$, following the study by Alexandrou *et al.* for static quarks[11]. In the present case, we found that n=5 gave statistically cleaner signals than n=0, but that n=10 was not an improvement over n=5.

In order to project out the mesonic ground states from the three-point function, one either needs excellent wave functions or large Euclidean time separations between the slices on which the heavy meson is created, emits a weak current, and is detected. Unfortunately, the statistical noise in the simulation overwhelms the signal when those separations are as large as 4 units of Euclidean time. The largest separations that provide a signal are

$$\begin{aligned} x_4 - y_4 &= 3 \\ y_4 - z_4 &= 3 \end{aligned} \quad (16)$$

At 2 units of Euclidean time separation, which corresponds to a greater contamination with higher mass states than at 3 units, the computed form factor is flatter. This presumably reflects Bjorken's sum rule, which states that the sum of the squares of the Isgur-Wise functions to all final states is $1$[12].

The results of the simulation together with a comparison to the data on $B \to D^* l \nu$ from both ARGUS and CLEO are shown in Figure 2. The agreement with the ARGUS and CLEO data with the simulation using 3 units $\Delta t$ of time separation is obviously rather spectacular. However, to know that such a results is reliable we must, at the very least, have available several values of $\Delta t$ and see that the result is stable at the largest values. Since there is a large change in the Isgur-Wise function between $\Delta t = 2$ and $\Delta t = 3$, and since the signal has degenerated into noise by $\Delta t = 4$, we simply have no way of knowing whether this agreement is significant or fortuitous.

A simple quadratic fit to the simulation results at $\Delta t = 3$ gives for the slope of the Isgur-Wise function at the origin:

$$\xi'(1) = -.95 \tag{17}$$

For the reasons just stated above, we regard the numerical results presented here as preliminary. There are corrections to apply and improvements to make, beyond the usual resort to larger ensembles of larger and finer lattices. In the present calculation we have neglected both the matching of the continuum HQET to QCD with quarks whose masses are far above the dynamical QCD scale, and the matching of the lattice HQET to the continuum version. One estimate of the latter is that it is a small effect[6], but it is necessary to study it systematically. A significant improvement would be to find operators for the *B* and *D* mesons that match the states very well. While the "Gaussian" smearing we have used improves the signal somewhat, far more systematic techniques have been developed for application to static quarks[13], and they should be extendable to lattice heavy quarks with finite classical velocity as well. A further enhancement would be the use of lattices generated with so-called improved actions, and propagators from which the leading finite lattice spacing errors have been removed.

**FIGURE CAPTIONS**

**Figure 1.** The heavy lines represent b and c quarks, the light one a u or d quark, and the wavy line the emission of a weak current.

**Figure 2.** The Universal Isgur-Wise Form Factor plotted vs the classical velocity product $v \cdot v'$. Five wave function smearing steps were used in each simulation. The top graph corresponds to 2 time steps each between the initial heavy meson source and the emission of the weak current and between the emission of the current and the final heavy meson sink. The middle graph corresponds to 3 time steps each, and the bottom one to 4 time steps each. The error bars on the simulation points are statistical only, and include averages over all configurations related by lattice symmetry. The solid lines are simply quadratic least squares fits to the simulations to guide the eye. The open squares are measurements from CLEO and the open diamonds are from ARGUS.

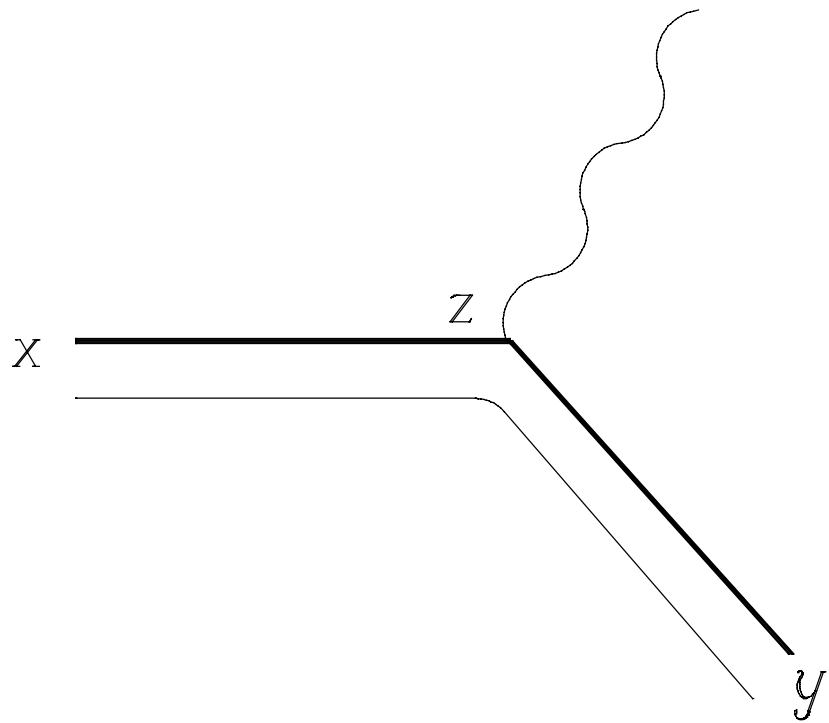

**Figure 1**

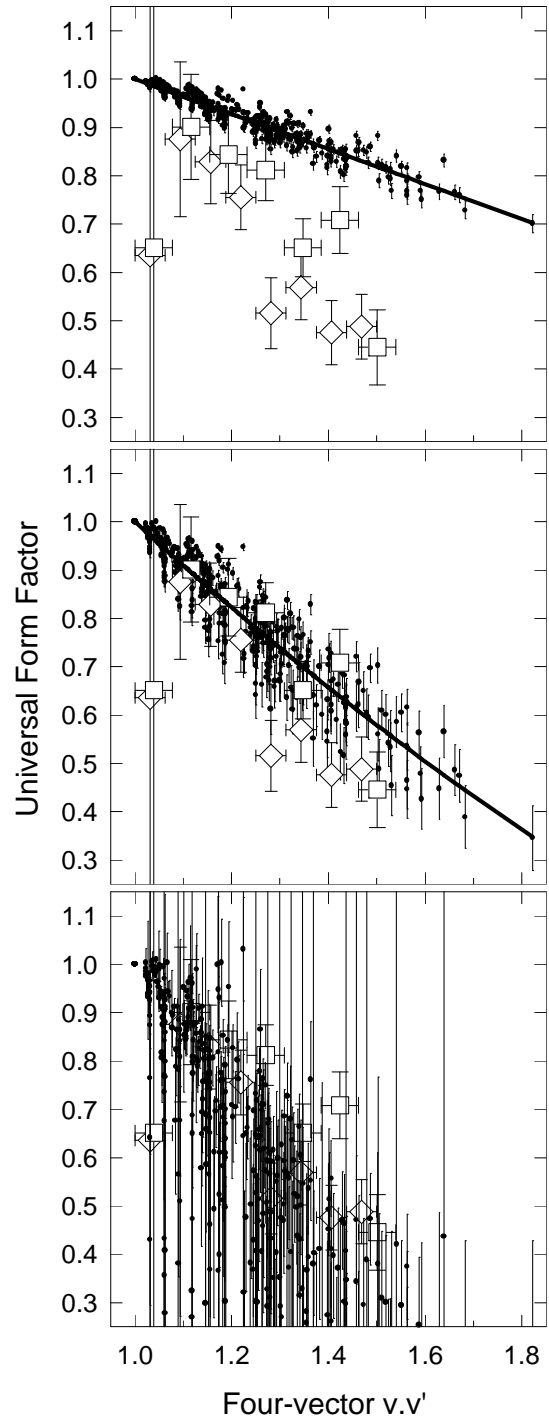

**Figure 2**